# A Bayesian Approach to Optimizing Stem Cell Cryopreservation protocols

S Sambu*

* Corresponding author:

P.O. Box 283,

Nandi Hills,

Kenya

Tel: +254 700 143225

Sambu@post.harvard.edu

## Abstract

Cryopreservation is beset with the challenge of protocol alignment across a wide range of cell types and process variables. By taking a cross-sectional assessment of previously published cryopreservation data (sample means and standard errors) as preliminary meta-data, a decision tree learning analysis (DTLA) was performed to develop an understanding of target survival and optimized pruning methods based on different approaches. Briefly, a clear direction on the decision process for selection of methods was developed with key choices being the cooling rate, plunge temperature on the one hand and biomaterial choice, use of composites (sugars and proteins), loading procedure and cell location in 3D scaffold on the other. Secondly, using machine learning and generalized approaches via the Naïve Bayes Classification (NBC) approach, these metadata were used to develop posterior probabilities for combinatorial approaches that were implicitly recorded in the metadata. These latter results showed that newer protocol choices developed using probability elicitation techniques can unearth improved protocols consistent with multiple unidimensional optimized physical protocols. In conclusion, this article proposes the use of DTLA models and subsequently NBC for the improvement of modern cryopreservation techniques through an integrative approach.

## 1. Introduction

Cryopreservation of cells often results in cell survivals that are lower because of suboptimal process variables. When cryopreservation is performed in biological constructs, the results are often worse due to the differentials in CPA concentration, CPA exposure times and cooling rates. These differences are exaggerated for cells that are farthest from the surface which is typical for large specimens(1). Conversely, small constructs that are just above the upper limit of the microcapsule diameter, differentials in heat loss are meagre(2). The main challenge for such constructs remains the varying CPA concentrations and exposure times for cells encapsulated in them. A spherical construct, for example, will have different cell survivals at each radius. The challenge is to minimize the deleterious effects of these differentials by providing cryoprotection without mass transfer limitations. Therefore, new construct variables arise. Hence, a re-prioritization of process and construct variables becomes necessary – a DLTA approach can help resolve the tensions in *prioritizing* and then *recursively* optimizing these variables.

In using a DTLA, it is necessary to leverage existing principles in cryopreservation. The entrapment of CPAs within the construct during cell resuspension in encapsulating material is an alternative strategy to the modulation of cell location. However, intracellular CPAs are often toxic and encapsulation is not lossless. Hence, when a potentially cytotoxic CPA such as DMSO is added to the encapsulating material, cell survival is often diminished(2)(3).Therefore, the success of the strategy relies on using a benign CPA as the “entrapped CPA” followed by the loading of the potentially malign but fast-diffusing CPA into the construct. For protein-based cryoprotection, mixing the proteins with polar CPAs also has a deleterious effect on the ability of the protein to confer cryoprotection due to the potential denaturation of the proteins prior to cooling (4). Overall, the protein-DMSO interaction is minimized by incorporating the protein in the encapsulating material, and allowing a time-limited diffusion of the malign CPA into the construct prior to cooling. Besides modifying the encapsulation process, slow-cooling procedures should also be adopted to minimize the required concentrations of the potentially malign CPA.

Besides using proteins and intracellular CPAs, synthetic non-penetrating polymers (SNPP) can also provide cryoprotection within the scaffold, thereby bypassing the limitations of diffusion in higher-dimensional cryopreservation. A subclass of these SNPPs are the vitrifying polymers which have been used to encourage extracellular vitrification of the cryopreservative during cooling thereby limiting ice crystal growth(5). Vinyl-derived polymers have also been shown to decrease ice crystal size colligatively. Examples include polyethylene glycol (PEG), polyvinyl alcohol (PVA) and hydroxyethyl starch. The key challenge in using these polymers is minimizing large increases in the viscosity of the solution (which will make encapsulation difficult) and minimizing the difficulty in post-thaw extraction of cells from the scaffold.

Shifting the focus from solutes to encapsulates, changing the encapsulation material (the matrix) properties by using polymer composites can be avoided if one entraps lower molecular weight solutes. One alternative would be to use low molecular weight polymers. Another would be to use sugars with cryoprotective properties. The mechanism for cryoprotection conferred by sugars is three-fold: complex sugars can inhibit the formation of ice(6,7); they can replace lost water molecules and can stabilize lipid bilayers from sudden phase transitions during cooling(8,9). They may prevent sudden changes in lipid phase and ensuing phase separation through hydrogen bond formation with phospholipid head groups(10,11). This interaction causes a reduction in the liquid crystal-gel transition temperature and minimizes membrane fusion by separating one layer from another(12). Sugars also stabilize proteins by encouraging a preferential hydration of the proteins in solution  (13). Granted that complex sugars are non-permeating solutes, these benefits are extracellular unless a delivery mechanism is engineered.

Each of the techniques mentioned can be represented as random decision variables (*factors*) in the abstract; each one taking on a probability state according to a protocol developed collectively and based on categorical field-specific knowledge. In that way, a DTLA will describe the epistemology of the factors and their relationship to a final probability state describing the life/death or living/apoptotic state of the hypothetical cell. However, the ability to generalize these DTL relationships is often severely reduced unless sophisticated pruning algorithms are employed(14) (conversely local results are definitively optimized based on the learning set) hence requiring the use of a more generalized approach taking the final hypothetical cell state as  discrete and each influencing factor as sufficiently independent in the decision-making process as well as in the execution of the protocol.

Once the requirement for broader applicability of a learned model is defined, a more general approach to making predictions on related but distinct data sets is required. In this latter case, the NBC presents a more efficient and direct approach to prediction. The assumption of factor independence shrinks the pre-prediction computational requirements significantly and speeds the overall time-to-decision making. Nonetheless, empirical evidence must be invoked to support a combined approach using both NBC and DTL(15). Overall, from the decision-maker's view, by combining the informative approach of the DTLA and the boundary-focussed NBC, the opportunity cost is low and the decision process is holistic and exhaustive in retrospect.

This paper seeks to demonstrate the promising strategies for successful cryopreservation of cells in 3D scaffolds by using a DTLA approach to develop a heuristic for approaching cryopreservation across many subjects. Given the cryopreservation challenge specifically concerns the opportunities in 3D cryopreservation; success is measured by a cell survival percentage that is either equal to or better than suspension cryopreservation.

## 2. Materials and Methods

### 2.1. Data Collection:

Data was collected and categorized into the main distinguishing categories namely: dimensionality, cell location, employment of a biomaterial, lyoprotection, CPA loading, CPA, containment, cooling rate and plunge temperature. Where original data was not present, the mean and standard error were used to redraw virtual samples from the original population described by the sample parameters.

### 2.2. Decision Tree Learning Analysis

Analysis was performed using *Rpart* package in the open source program R for recursive partitioning and regression analysis.

In brief, the collected meta-data(16,17),(18) was imported into R, partitioned according to a variable that best describes the data at each given partition level. Thereafter, the model is cross-validated and the model that best describes the data set is selected after evaluation for fidelity to the original data.

### 2.3. Naïve Bayes Classification (NBC)

The *e1071* package[17] was used to run the NBC process using the decision factors of dimensionality, location, biomaterial, sugars, step-loading, integrins with the terminal decision variable as cell survival represented in two independent states: apoptotic or live.

## 3. Results and Discussion

### 3.1. A DTL analysis focussed on material and process choices

From Figure 1, processes involving the incorporation of sugars improve survival as is captured by the partition labelled *"Sugars=a"* with a 7% delta across both predictions. The incorporation of sucrose into hydrophilic biomaterials modulates porosity e.g. in the alleviation of decreased porosity caused by material compression(13). In the context of cryopreservation, the addition of sucrose helped create further pore interconnections in the final scaffold and/or create new voids. This is primarily because of the inhibition of DMSO action by van der Waals (VWF) and hydrogen-bonding interactions between sugars, water and DMSO. These sugar-DMSO and water-DMSO interactions are expected because DMSO is prone to act as a hydrogen bond acceptor with both sugar and water molecules[14]. The addition of sucrose as a component of the alginate capsule and as part of the cryopreservative improves the post-thaw cell viability as previous studies of such systems have shown that the effective concentration of water and the fraction of freezable water are diminished for alginate concentrations greater that 0.5% (w/w)(19). This is consistent with diminished ice formation and freeze-concentration. The addition of sugars to the alginate scaffold was thought to be necessary since mass transfer barriers are eliminated and the cryoprotective benefits of sugar were expected to lower the required minimum DMSO exposure times and intracellular DMSO content(20). Whether in the cryopreservative or in the scaffold, the mechanism for cryoprotection involves hydrogen bonding between atoms within hydroxyl group of the sugar, DMSO and water. Sucrose can act as a hydrogen bond donor to both DMSO and water. DMSO is known to isolate water molecules and to induce hydrogen bonding(21). This may be related to the interaction between trehalose and alginate solutions; it has already been established that sugars impact the swelling ratios of hydrogels and by the same token, the survival of encapsulated cells(22). The higher swelling ratio of the trehalose-alginate system is linked to the lower impact trehalose has on the effective concentration of water available for alginate swelling. This will have an impact in the cryopreservation of cells since the relatively higher effective concentration of water in the alginate-trehalose system will increase the probability for ice formation which is lethal to cells.

Overall, it is expected that two measurable quantities i.e. the fraction of freezable water and the effective concentration of water in alginate/trehalose scaffolds will be higher(23); this will in turn mean that the level of cryoprotection in these scaffolds will be lower than that observed for sucrose consistent with experimental observations.

Given the differentials in cell survival for 3D constructs, the provision of cryoprotection throughout the scaffold has been shown to improve overall cell survival and to minimize differentials along the construct radius; this will ensure that long exposure times are eliminated from the cryopreservation process(24). Comparable results indicate that cryopreservation requirements place even greater diffusion limits than do nutrient and oxygen perfusion limits (< 2mm); this is corroborated by the 9% (see *"location=b"* in Figure 1)increase caused by the "location" of cells within the 3D construct.

From Figure 1, there are competitive alternatives in the neighbourhood of the use of scaffolds with biomechanical signalling, namely:-

1. The use of sugars, natural biomaterials and biomechanical signalling via integrins (87%)
2. Sugars and natural biomaterials (84%)
3. Sugars, natural biomaterials and location of cells in shell (82%)
4. Sugars and natural biomaterials (79%)

All four decisions are from the right-most termini of the decision tree and are optimal in their own respect; however, overall, the employment of integrins is by far the best approach.

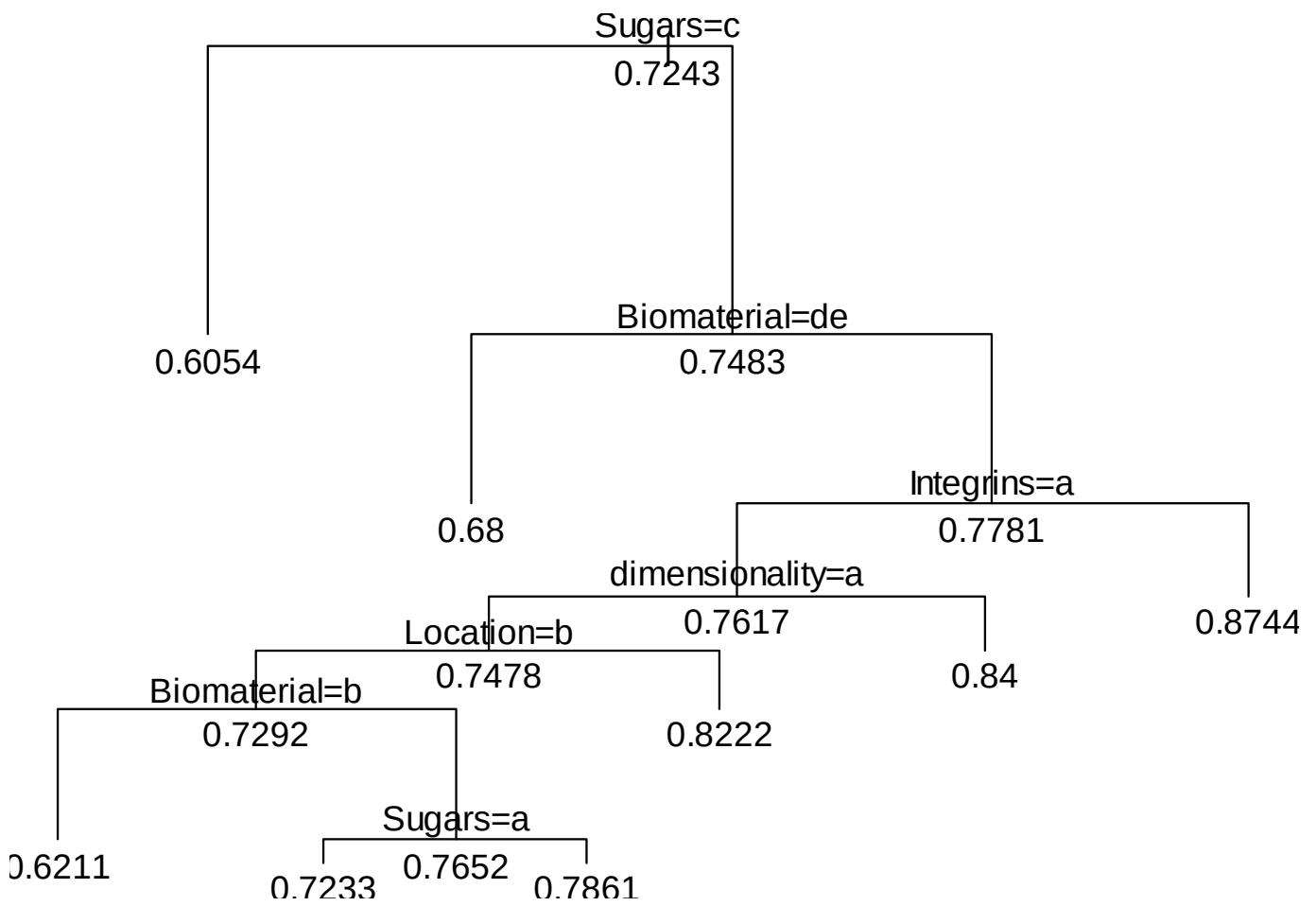

**Figure 1: A recursive partitioning in detail showing the hot-spots at "Integrins", "Sucrose/Steploading(Two)" both of which employ "Biomaterial" during cryopreservation.**

From Figure 2, the DLT approach is locally optimal as confirmed by the $R^2$ of 1 between predicted and actual showing the prediction captures all the variability inherent in the metadata.

However, there is a limit to the DTLA approach in that generalization cannot be delivered as there are no predictions outside the very local and very limited scope.

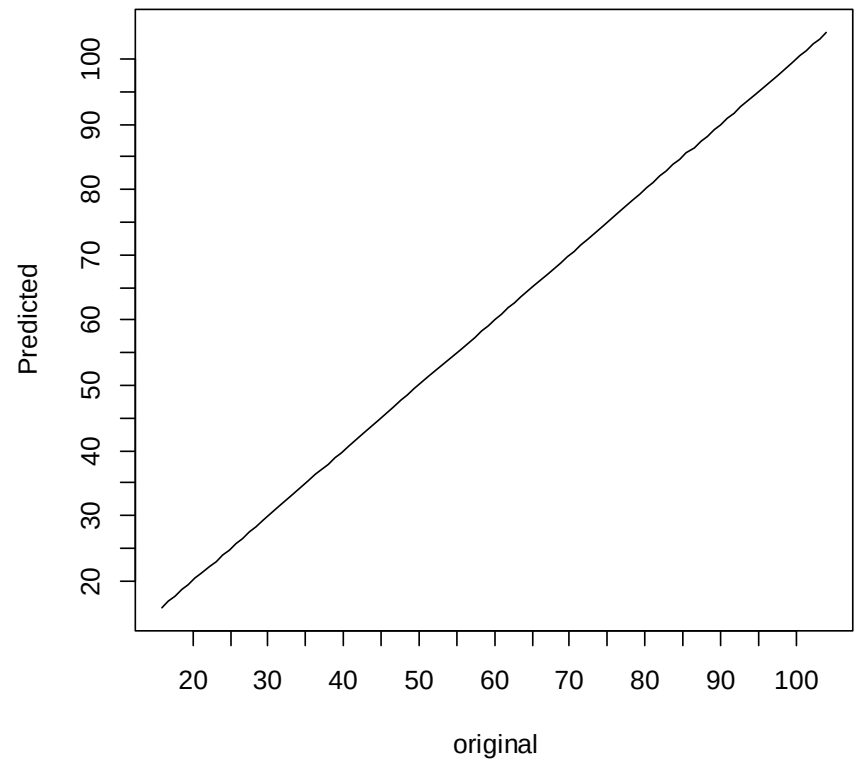

**Figure 2: A plot of the predicted against the original data for the decision tree learning showing a a perfect fit based on the local optimization of each decision terminus as relates to the factors**

## 3.2. An augmented DTL analysis focussed on thermal history, material and process choices

Further analysis with augmented data(18) via the DTLA enabled further refinement of the decision process per Figure 3 which shows that the local maximum is sustained against new process variables related to thermal history i.e. cooling rates and plunge temperatures. Given the branching mode, it is clear that process variables lead to a different decision-making path than the matrix related variables. Of note, changes on process variables lead to significant changes in the survival (see 18% increase on *Plunge Temperature < -60.5°C).* In Figure 3, a meta-data analysis shows a survival-optimized/conserved right-side analysis whereby the ranking for protocol decisions were as follows:-

1. Sugars, natural biomaterials and biomechanical signalling via integrins (87%)
2. Sugars, vitrifying extracellular polymers and in suspension (84%)
3. Sugars and natural biomaterials (78%)
4. Sugars, natural biomaterials and location of cells in outer core (78%)

However, the limit in this heuristic is in generalization i.e. no predictions can be made outside the local and very limited original scope of the specific combinations within the meta-data.

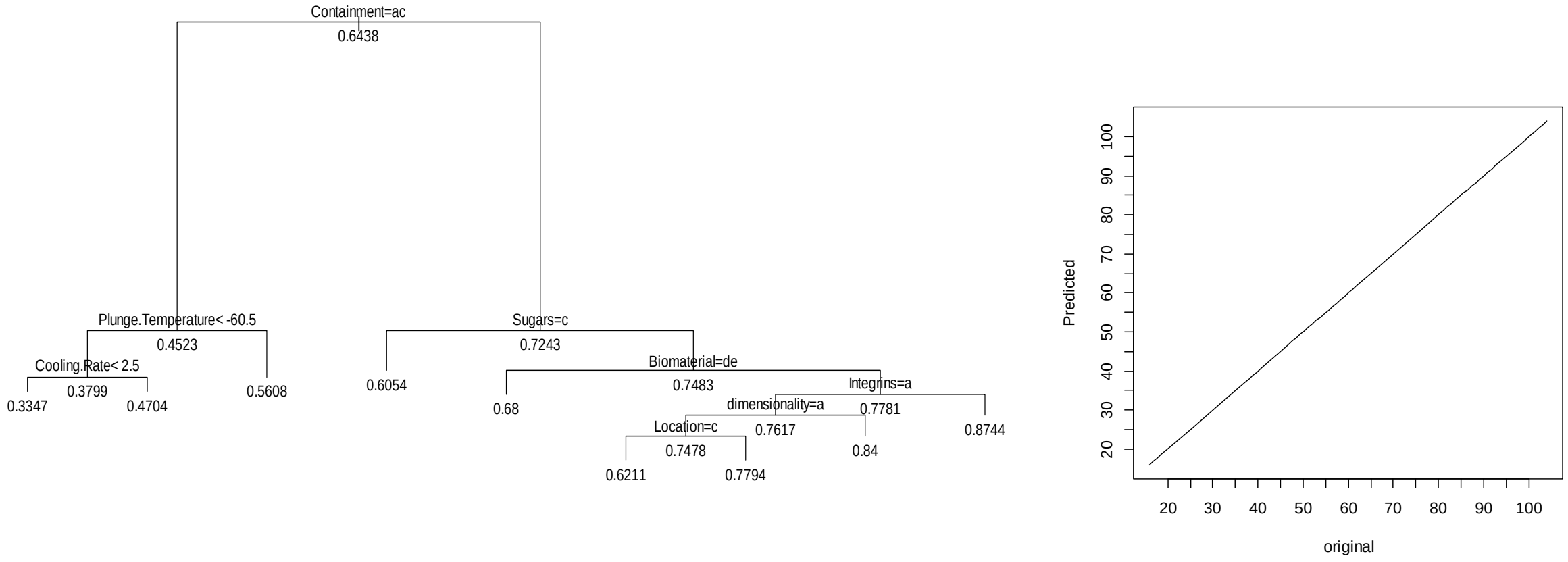

**Figure 3: A meta-data analysis of cryopreservation data showing that the right-side analysis is conserved across methods; (R) plot showing agreement between the original data subset and the metadata subset**

Before a generalization can be developed, an examination of patterns to follow in the construction of the generalizable classifier is required -- Figure 4 captures the scatter plot matrix of the meta-data to be used in generating the generalizable classifier. From the matrix, the graphs that have all green dots are outlined in purple. The variables i.e. *Step-loading*, *integrins* and *location* are particularly pivotal in the attainment of high post-thaw cell survival rates. One additional factor, i.e. *sugars*, when intersected with *location* also provided one of the 7 graphs with exceptional survival rates.

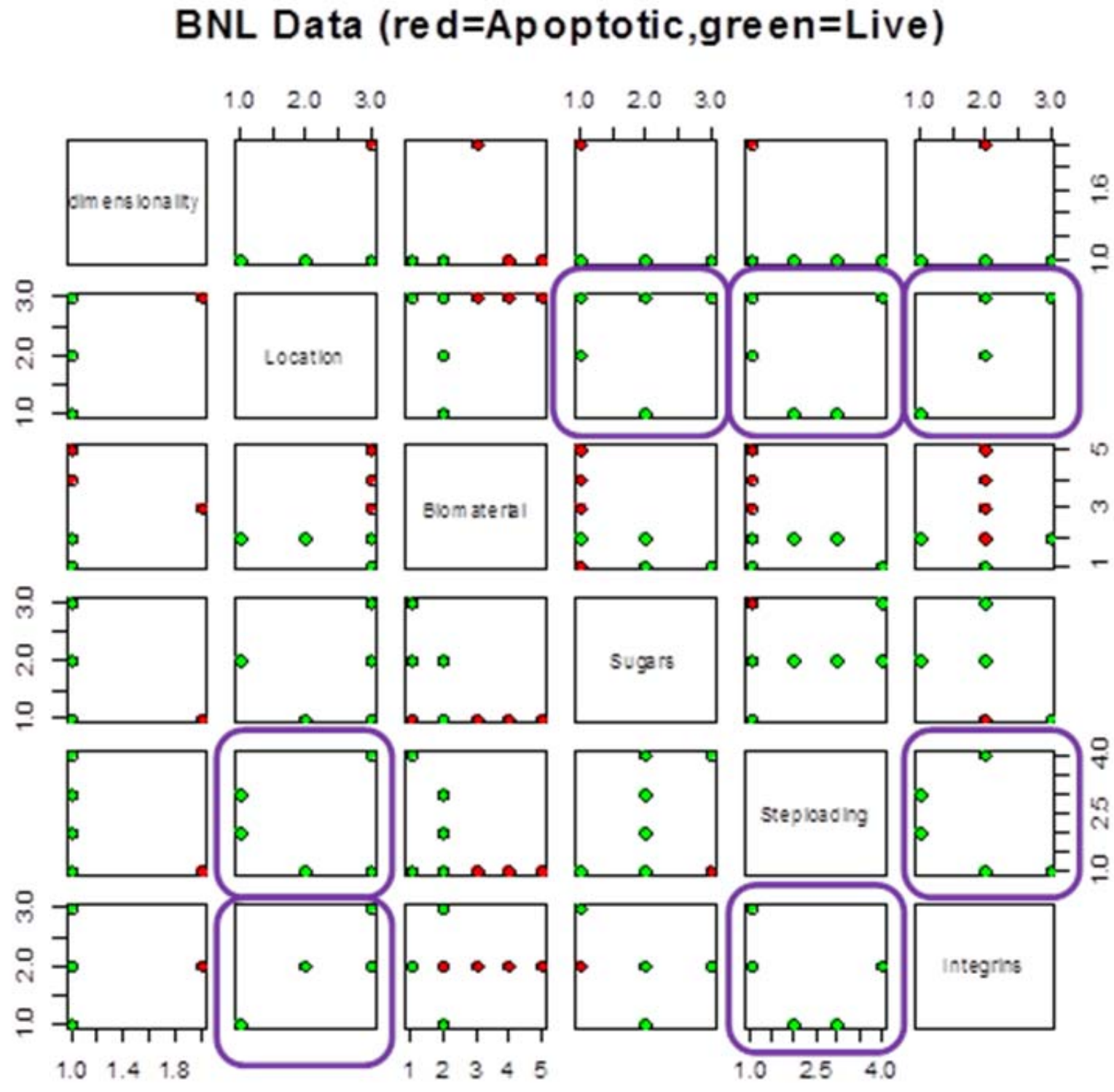

**Figure 4: A scatter plot matrix capturing the meta-data collected from previous analyses used to develop a heuristic for predicting the posterior survival probabilities for cells for a given set of process decisions(17,18,25). Key decision variable combinations with 100% *"live"* cells are framed in purple.**

Figure 5 shows how a change in survival against the *plunge temperature* set point significantly changes the terminal/'leaf' value. This is the temperature at which the cryovials will be immersed in liquid nitrogen. From Figure 5, the plunge temperature set point is at -60 $^{o}$C -- while this may be applicable to the specific protocols used(18), there may be difficulties in generalization outside of the cell type used unless the models developed are able to learn across this data set.

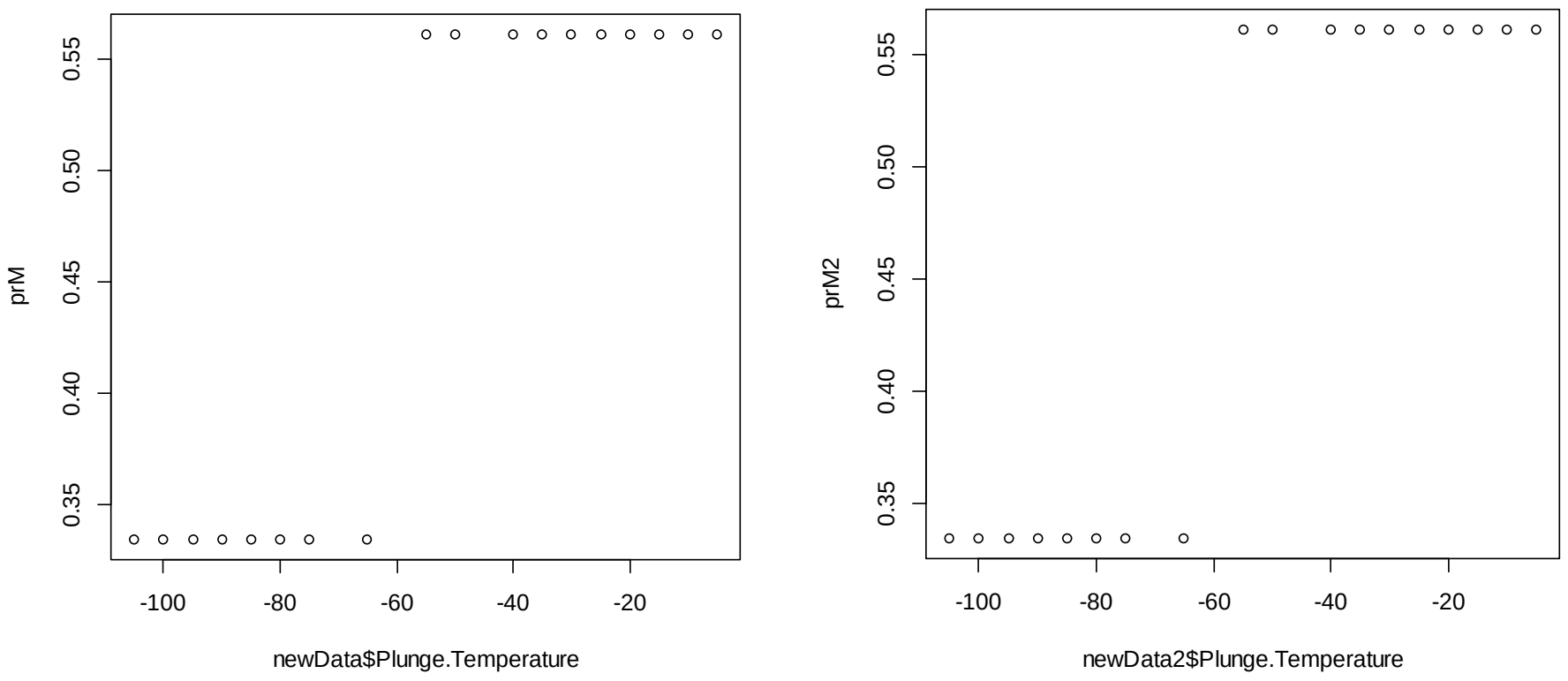

**Figure 5: Change in survival against temperature capturing the temperature set point against which the prediction changes the 'leaf' value. A 2D plot of the survival rate against temperature was drawn from a DTL analysis**

Figure 6 shows results from the NBC method which allows generalization across the factors used whereby new combinations of otherwise independent decision factors can be made and evaluated for posterior probabilities. These predictive syntheses demonstrate a generalized accuracy of 79% across the meta-data while the accuracy for predictive analytics for a specified case whereby samples are all cryopreserved in 3D natural RGD-containing matrices, using controlled slow-cooling with sucrose for lyoprotection, in a two-step loading process shows an 89% accuracy for a survival of 89% of original viable cells which meets the needs for evaluation of most-likely protocol choices. After these results were derived, an analysis of recent published and peer-reviewed data showed that loss of RGD-responsive cell attachments during cryopreservation is a leading cause of cell loss(26)further, that sucrose as a porogen helps provide *in vivo*-like conditions in a step-wise process during tissue fabrication(27). Subsequent to these initial results, further work has shown that expansion and cryopreservation in 3D microcapsule matrices for cellular aggregates is demonstrable for human embryonic stem cells further confirming that decision choices that encompass cell attachment (similar in function to RGD-containing matrices), slow-cooling, matrix permeability, natural biomaterials (e.g. alginate) will enhance cell viability(28).

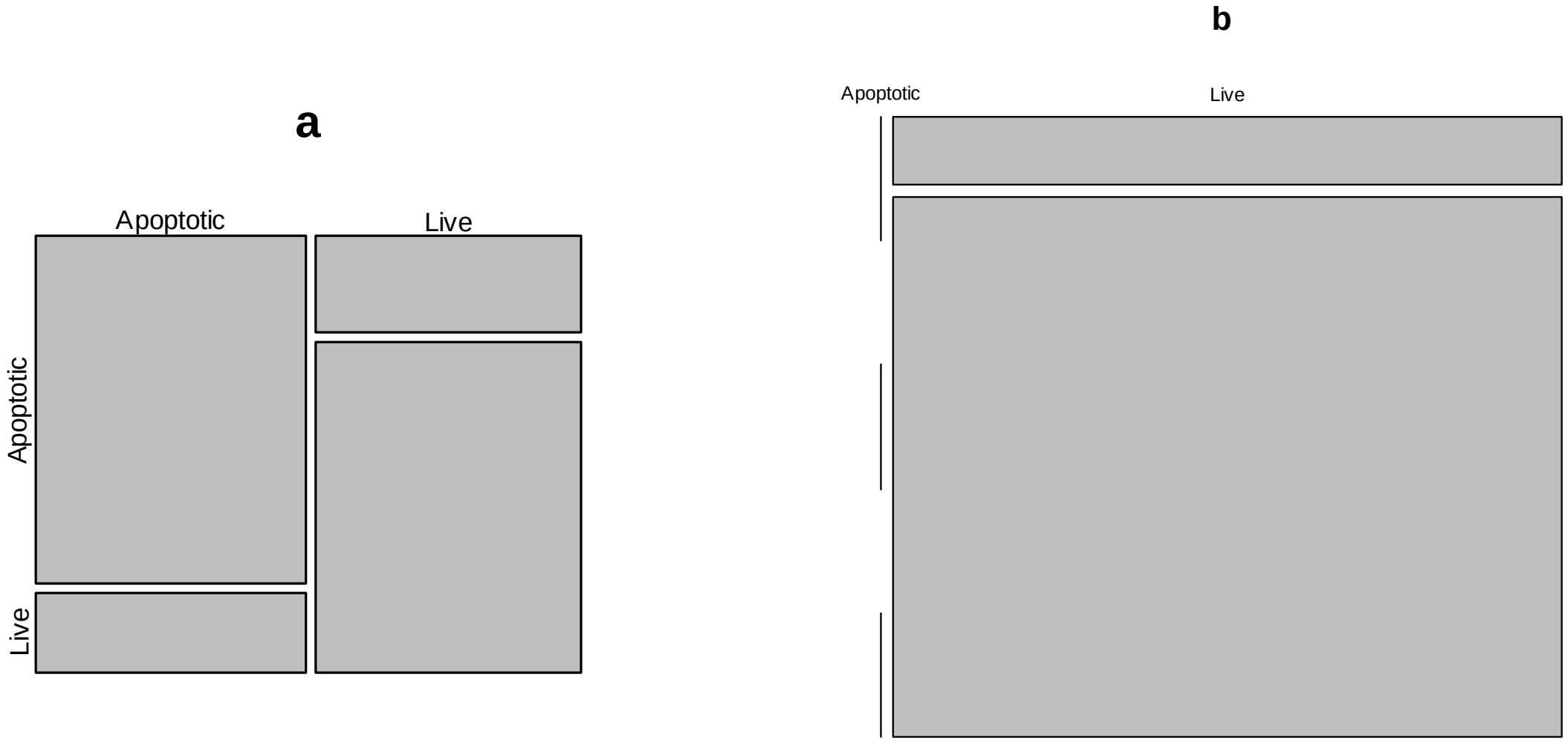

**Figure 6: Prediction of survival using the Naïve Bayes Classifier on the following factors: dimensionality, location, biomaterial, sugars, step-loading & integrins. Graph "a" shows predictions for the training set with 79% accuracy while the test set shows an 89% accuracy showing that the Naïve Bayes Classifier is a demonstrably accurate predictor when reduced to a smaller test set where the samples are all cryopreserved in 3D natural RGD-containing matrices, using controlled slow-cooling with sucrose for lyoprotection, in a two-step loading process [Similar to the right-most branch of the recursive partitioning trees developed earlier]**

### 3.3. Synthesis

A synthesis of the previous analyses shows that the use integrin ligands improve cell survival sharply. However, where cost may be a limiting factor, the use of sugars and a careful biomaterial choice may suffice as alternatives confirmed by recursive partitioning. This result is definitively captured by the decision tree model which shows that the lead attributes for cryopreservation are: sugars, integrins, dimensionality and location according to the information gain these attributes afford at each cycle of recursive testing.

## 4. Conclusion

In conclusion, a DTL was employed on mESC cryopreservation meta-data. These results indicate that the handles for improving cryopreservation outcomes are: integrin-mediated cryopreservation, the modification (by entrapment of benign CPAs or other means) of the scaffolding material and the modification of cell location in scaffolds. The DTL model was demonstrated to be robust against tougher validation data from a different cell types thereby confirming that cryopreservation/bio-preservation technology can be improved upon using a DTL approach.

## Bibliography

1. Muldrew K, Novak K, Yang H, Zernicke R, Schachar NS, McGann LE. Cryobiology of Articular Cartilage: Ice Morphology and Recovery of Chondrocytes. Cryobiology. 2000 Mar;40(2):102–9.

2. Cui ZF, Dykhuizen RC, Nerem RM, Sembanis A. Modeling of Cryopreservation of Engineered Tissues with One-Dimensional Geometry. Biotechnol Prog. 2002;18(2):354–61.

3. Liu Y. Cryopreservation of Mouse embryonic Stem Cells. 2007.

4. Meryman HT. Cryopreservation of living cells: principles and practice. Transfusion (Paris). 2007;47(5):935–45.

5. Gibson MI, Barker CA, Spain SG, Albertin L, Cameron NR. Inhibition of Ice Crystal Growth by Synthetic Glycopolymers: Implications for the Rational Design of Antifreeze Glycoprotein Mimics. Biomacromolecules. 2009;10(2):328–33.

6. Roos Y, Karel M. Amorphous state and delayed ice formation in sucrose solutions. Int J Food Sci Technol. 1991;26(6):553–66.

7. Shirakashi R, Müller KJ, Sukhorukov VL, and Zimmermann U. Effects of physiological isotonic cryoprotectants on living cells during the freezing-thawing process and effects of their uptake by electroporation: Sp 2 cells in alginate-trehalose solutions. Heat Transf Asian Res. 2003;32(6):511–23.

8. Arakawa T, Carpenter JF, Kita YA, Crowe JH. The basis for toxicity of certain cryoprotectants: a hypothesis. Cryobiology. 1990;27(1):401–15.

9. Rudolph AS, Crowe JH. Membrane stabilization during freezing: the role of two natural cryoprotectants, trehalose and proline. Cryobiology. 1985;22(4):367–77.

10. Prestrelski SJ, Arakawa T, Carpenter JF. Separation of freezing- and drying-induced denaturation of lyophilized proteins using stress-specific stabilization. II. Structural studies using infrared spectroscopy. Arch Biochem Biophys. 1993;303(2):465–73.

11. Anchordoguy TJ, Rudolph AS, Carpenter JF, Crowe JH. Modes of interaction of cryoprotectants with membrane phospholipids during freezing. Cryobiology. 1987;24(4):324–31.

12. Arakawa T, Timasheff SN. Stabilization of protein structure by sugars. Biochemistry (Mosc). 1982;21(25):6536–44.

13. Huang YC, Connell M, Park Y, Mooney DJ, Rice KG. Fabrication and in vitro testing of polymeric delivery system for condensed DNA. J Biomed Mater Res 67A 4. 2003;4(Journal Article):1384–92.

14. Fürnkranz J. Pruning Algorithms for Rule Learning. Mach Learn. 1997 May 1;27(2):139–72.

15. Rubinstein YD, Hastie T. Discriminative vs Informative Learning. KDD. 1997. p. 49–53.

16. Sambu S, Xu X, Schiffter H, Cui Z, Ye H. RGDS-Fuctionalized Alginates Improve the Survival Rate of Encapsulated Embryonic Stem Cells During Cryopreservation. Cryoletters. 2011 Sep 1;32(5):389–401.

17. Heng BC, Yu Y-JH, Ng SC. Slow-cooling protocols for microcapsule cryopreservation. J Microencapsul. 2004;21(4):455–67.

18. Kashuba Benson CM, Benson JD, Critser JK. An improved cryopreservation method for a mouse embryonic stem cell line. Cryobiology. 2008;56(2):120–30.

19. Pongsawatmanit R, Ikeda S, Miyawaki O. Effect of sucrose on physical properties of alginate dispersed aqueous systems. Food Sci Technol Res. 1999;5(2):183–7.

20. MacGregor WS. The chemical and physical properties of DMSO. Ann N Y Acad Sci. 1967;141(Biological Actions of Dimethyl Sulfoxide):3–12.

21. Sato Y, Kawabuchi S, Irimoto Y, Miyawaki O. Effect of water activity and solvent-ordering on intermolecular interaction of high-methoxyl pectins in various sugar solutions. Food Hydrocoll. 2004;18(4):527–34.

22. Selmer-Olsen E, Sørhaug T, Birkeland SE, Pehrson R. Survival of Lactobacillus helveticus entrapped in Ca-alginate in relation to water content, storage and rehydration. J Ind Microbiol Biotechnol. 1999;23(2):79–85.

23. Martinsen A, Storrø I, Skjårk-Bræk G. Alginate as immobilization material: III. Diffusional properties. Biotechnol Bioeng. 1992;39(2):186–94.

24. Griffith CK, Miller C, Sainson RCA, Calvert JW, Jeon NL, Hughes CCW, et al. Diffusion limits of an in vitro thick prevascularized tissue. Tissue Eng. 2005;11(1-2):257–66.

25. Miszta-Lane H, Gill P, Mirbolooki M, Lakey JRT. Effect of Slow Freezing Versus Vitrification on the Recovery of Mouse Embryonic Stem Cells. Cell Preserv Technol. 2007;5(1):16–24.

26. Terry C, Hughes RD, Mitry RR, Lehec SC, Dhawan A. Cryopreservation-Induced Nonattachment of Human Hepatocytes: Role of Adhesion Molecules. Cell Transplant. 2007 Jun 1;16(6):639–47.

27. Verhulsel M, Vignes M, Descroix S, Malaquin L, Vignjevic DM, Viovy J-L. A review of microfabrication and hydrogel engineering for micro-organs on chips. Biomaterials. 2014 Feb;35(6):1816–32.

28. Serra M, Correia C, Malpique R, Brito C, Jensen J, Bjorquist P, et al. Microencapsulation Technology: A Powerful Tool for Integrating Expansion and Cryopreservation of Human Embryonic Stem Cells. PLoS ONE. 2011 Aug 5;6(8):e23212.